# Impact of shadow banks on financial contagion

Yoshiharu Maeno[1], Kenji Nishiguchi[2], Satoshi Morinaga[1], Hirokazu Matsushima[3]

[1]NEC Corporation, Kawasaki, Japan

y-maeno@aj.jp.nec.com

[2]Japan Research Institute, Tokyo, Japan

[3]Institute for International Socio-economic Studies, Tokyo, Japan

**Abstract.** An asset network systemic risk (ANWSER) model is presented to investigate the impact of how shadow banks are intermingled in a financial system on the severity of financial contagion. Particularly, the focus of this study is the impact of the following three representative topologies of an interbank loan network between shadow banks and regulated banks. (1) Random mixing network: shadow banks and regulated banks are intermingled randomly. (2) Asset-correlated mixing network: banks having bigger assets are a regulated bank and other banks are shadow banks. (3) Layered mixing network: banks in a shadow bank layer are connected to banks in a regulated bank layer with some interbank loans.

## 1 Introduction

Understanding how the characteristics of a financial system govern the financial contagion of bank bankruptcies is essential in the argument to reform the capital requirement and other regulatory standards. Recently computer simulation models [10], [12], [13], [16] are developed to mimic the transmission of financial distress and predict the systemic risk [3], [6], which is the severity of financial contagion in a financial crisis. Both the external assets and interbank loans of banks can be the origin of financial distress in these models. Either distress may transmit separately in a peace time while compound distress transmits in a crisis time. A bank makes an investment in multiple external asset classes. The value of the total external assets may depreciate when the markets fluctuate. A defective investment portfolio of banks imposes financial distress on them. A failing debtor bank becomes insolvent in paying off the interbank borrowings. Any creditor banks suffer financial distress from the failing debtor bank. A bank goes bankrupt unless the capital buffer absorbs the total loss from the external assets and interbank loans. Bank bankruptcies bring about still more financial distress repeatedly. This is the mechanism of financial contagion.

A shadow banking system [2], [4], [5], [15] had grown rapidly to rival depository banks after 2000. A shadow bank is not a regulated bank, but such an unregulated financial intermediary as hedge funds, money market mutual funds, and investment banks. It is not subject to the international Basel III requirements on capital buffer, and monitoring by government authorities. Shadow banks have a very high level of leverage, which is a very high ratio of debts to liquid assets, through off-balance sheet financing. Consequently, they have merely a very poor capital buffer. The vulnerability of the shadow banking system was one of primary factors to cause the global financial crisis in 2008 that ensued from the collapse of the US subprime mortgage markets. The US government authorities placed significant blame for the freezing of credit markets on a run on shadow banks, which had borrowed short-term in liquid markets to purchase illiquid risky assets.

In this study, we present a computer simulation model, which is called an asset network systemic risk (ANWSER) model [1], [8], [9], [11], to investigate the impact of how shadow banks are intermingled in a financial system on the severity of financial contagion. Particularly, we are interested in the impact of the following three representative topologies of an interbank loan network between shadow banks and regulated banks.

(1) In the random mixing network, shadow banks and regulated banks are intermingled randomly. It is a reference network topology. (2) In the asset-correlated mixing network, banks having bigger assets are regulated banks and other banks are shadow banks. It is less trivial than the reference network topology. Big banks are subject to regulatory standards worldwide. It was announced in 2011 that the international Basel III requirements would impose a relatively high level of capital buffer, which means additional loss absorbency, on global systemically important banks. The fraction of shadow banks may have a big impact on financial contagion. (3) In the layered mixing network, a financial system consists of two bank layers, and banks in the shadow bank layer are connected to banks in a regulated bank layer with interbank loans. It is even less trivial than the reference network topology. The number of inter-layer loan relations may have a big impact on financial contagion.

## 2 ANWSER model

Models of interbank loans and investments are presented in this section.

The asset network systemic risk model (ANWSER) is founded on previous computer simulation models. They investigate the statistical characteristics of a financial system with a Monte-Carlo method. The Monte-Carlo method is a broad class of a computational technique to obtain many samples of numerical outcomes which are used to analyze the statistical characteristics. The technique relies on a sequence of random numbers generated repeatedly from a specified probability distribution. The initial financial distress on banks is the falling prices of their external assets in the market. When a debtor bank happens to go bankrupt, the consequent interbank loan defaults are the next financial distress to its creditor banks. Financial distress transmits from failing debtor banks to creditor banks repeatedly in an interbank network.

The number of banks is $N$. $M$ is the number of external asset classes in which an individual bank makes an investment. The interbank loan ratio of a financial system $\theta = \Sigma l_n / \Sigma a_n$ is the total value of interbank loans as a fraction of the total value of assets. The assets of the $n$-th bank consist of the interbank loans $l_n$ and external assets $e_n$. The external assets are securities and government bonds. An interbank loan is the credit relation between a creditor bank and a debtor bank which appears when the debtor bank raises money in the interbank market. An interbank network describes the all credit relations. It is a directed graph which consists of banks as vertices, and the interbank loans as edges from creditor banks to debtor banks. The liability consists of the equity capital $c_n$, interbank borrowings $b_n$, and deposits $d_n$. The equity capital includes common stock and disclosed reserves. The equity capital ratio (core tier 1 ratio) is $\gamma_n = \Sigma c_n / \Sigma a_n$. These need not be paid off and can be used to absorb the loss from financial distress immediately. The amount of the assets is equal to that of the liability in the balance sheet, $a_n = l_n + e_n = c_n + b_n + d_n$.

The denseness $\kappa$ of a financial system is the average incoming or outgoing nodal degree of the interbank network as a fraction of $N - 1$. A more dense interbank network has a larger value of $\kappa$. The concentration $\rho$ of a financial system is the sum of the interbank loan share of the five biggest banks. A more concentrated interbank network has a larger value of $\rho$.

Given $N$ and $M$, a sequence of random numbers is generated to synthesize a number samples for fixed values of $\theta$, $\gamma_n$, $\kappa$, and $\rho$. An individual sample includes:
(1) Interbank network topology $\mathbf{Z}$ (an $N \times N$ matrix) where the element $Z_{nn'} = 1$ means the $n'$-th bank makes a loan from the $n$-th bank, and otherwise $Z_{nn'} = 0$.

(2) Investment portfolio $X$ (an $N \times M$ matrix) where the element $X_{nm}$ is the fraction of the investment which the $n$-th bank makes in the $m$-th external asset class ($\Sigma X_{nm} = 1, \ 0 \leq X_{nm} \leq 1$).
(3) Prices of the external assets in the market $\boldsymbol{v}$ (an $M$ column vector) where the element $v_m$ is the price of the unit of the $m$-th external asset class.

The initial financial distress on the $n$-th bank is $e_n \Sigma X_{nm} v_m$. It is assumed that failing debtor banks do not pay off any portions of the interbank loans to creditor banks. A bank goes bankrupt if the total loss from the financial distress is not absorbed by its capital buffer $c_n$. $F$ is the number of banks which end in bankruptcy until the financial contagion comes to a halt. The empirical distribution of the number of bank bankruptcies $P(F)$ is obtained from those samples. The value of $F$ is picked up at the 999-th 1000-quantile point as the representative in case of a financial crisis.

It is known empirically that the nodal degree of the network and the value of the transferred funds between banks obey a power law. In this study, $\boldsymbol{Z}$ is generated randomly by a generalized Barabasi-Albert model. This is a random graph with the mechanism of growth and preferential attachment which becomes scale-free as $N$ goes to infinity. The distribution of the nodal degree $k$ obeys the power law $P(k) \propto k^{-\alpha}$ where $\alpha \geq 2$. There is a significant probability of the presence of very big banks. This is the origin of heterogeneity.

The value of a loan $w_{nn'}$ from the $n$-the bank to the $n'$-th bank is determined from the incoming nodal degree $k_n^{in}$ and outgoing nodal degree $k_{n'}^{out}$ in the interbank network topology by the generalized law: $w_{nn'} \propto \left(k_n^{in} k_{n'}^{out}\right)^r$. The concentration $\rho$ increases as $r \geq 0$ increases. The value of interbank loans is a constant if $r = 0$. Once the value of $w_{nn'}$ is given, the interbank loans and borrowings of individual banks are determined. Then the balance sheet of individual banks is determined from the values of $\theta$ and $\gamma_n$. A prerequisite that the external assets are no less than the net interbank borrowings are imposed because the bank has already gone bankrupt if this prerequisite is not satisfied.

A bank chooses multiple external asset classes to make an investment in randomly. When $M = 2$, $X_{n1}$ and $X_{n2}$ obey a uniform distribution. The prices of the external asset classes are independently and identically distributed. The absolute fluctuation in their prices obeys a uni-variate Student $t$-distribution. The prices rise or fall randomly. The degree of freedom is $\mu = 1.5$. This is a long tailed distribution which is suitable to describe a sudden large fluctuation. The amplitude of the absolute fluctuation is adjusted so that the probability of a bank with the equity capital ratio $\gamma = 0.07$ alone going bankrupt can be $p = 10^{-3}$.

## 3   Network topology

The focus of this study is the impact of the following three representative topologies of an interbank loan network between shadow banks and regulated banks. Empirically, some of real financial system may be close to the asset-correlated mixing network. Some of government authorities seem to believe the layered mixing network is relatively robust.

(1) Random mixing network: Shadow banks and regulated banks are intermingled randomly. It is a reference network topology. The number of shadow bank as a fraction of $N$ is $0 \leq f \leq 1$. Fig. 1 shows an example topology when $N = 30, f=0.5$. Blue nodes are shadow banks. Their number is $N_s = 0.5N$. Red nodes are regulated banks. Their number is $N_r = N - N_s = 0.5N$.

(2) Asset-correlated mixing network: Banks having bigger assets are regulated banks and other banks are shadow banks. It is less trivial than the reference network topology. Big banks are subject to regulatory standards worldwide. It was announced in 2011 that the international Basel III requirements would impose a relatively high level of capital buffer, which means additional loss absorbency, on global systemically important banks. As a result, the equity capital ratio is correlated to the amount of assets generally in a real financial system. Such a core-periphery network as the asset-correlated mixing network is a simplified but still substantial replica of the real financial network. The number of shadow bank as a fraction of $N$ is $0 \leq f \leq 1$. The fraction of shadow banks may have a big impact on financial contagion. Fig. 2 shows an example topology. Shadow banks are peripheral small banks. Regulated banks are central big banks.

(3) Layered mixing network: A financial system consists of two bank layers, and banks in the shadow bank layer are connected to banks in a regulated bank layer with interbank loans. It is even less trivial than the reference network topology. Some of government authorities believe separating a financial system into multiple bank layers is effective in mitigating the severity of financial contagion. The layered mixing network is reasonable in such a regulatory belief. The number of shadow bank as a fraction of $N$ is $f = 0.5$. The number of inter-layer loan relations may have a big impact on financial contagion. The denseness of the inter-layer links is $q\kappa$, while the denseness in the intra-layer links is $\kappa$. The quantity $0 \leq q \leq 1$ adjusts the inter-layer denseness relative to the intra-layer denseness. The two bank layers are decoupled completely when $q = 0$. The denseness is uniform all over the network when $q = 1$. Fig. 3 shows an example topology. A non-trivial layer structure is visible clearly.

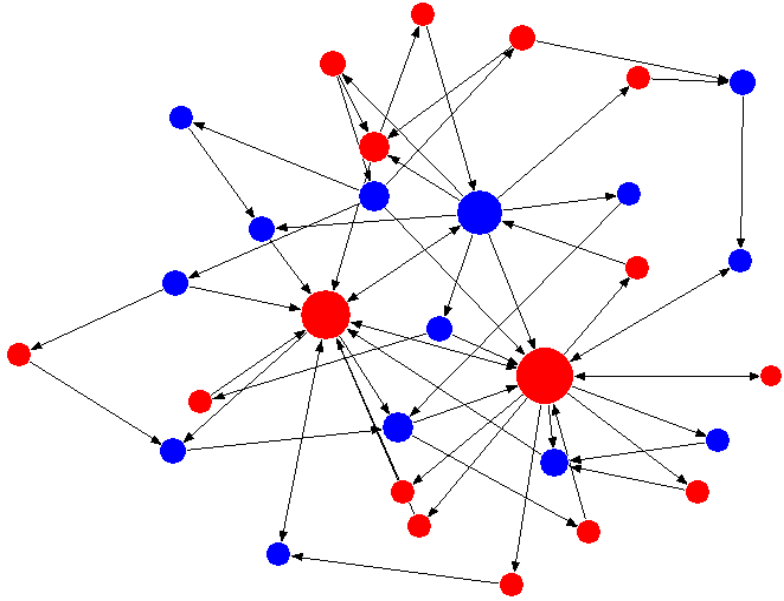

**Fig. 1** Example topology of a random mixing network when $N_r = N_s = 0.5N, N = 30$. Blue nodes are shadow banks. Red nodes are regulated banks.

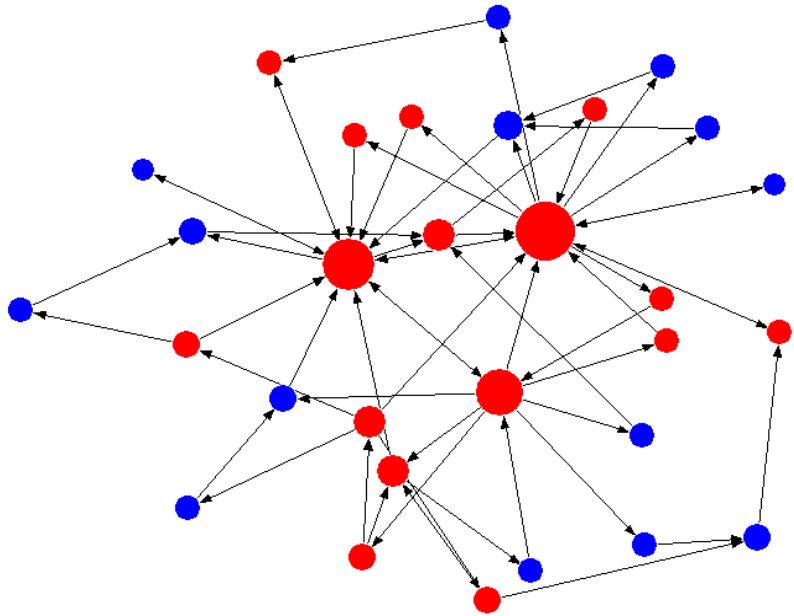

**Fig. 2** Example topology of an asset-correlated mixing network when $N_r = N_s = 0.5N, N = 30$. Blue nodes are shadow banks. Red nodes are regulated banks.

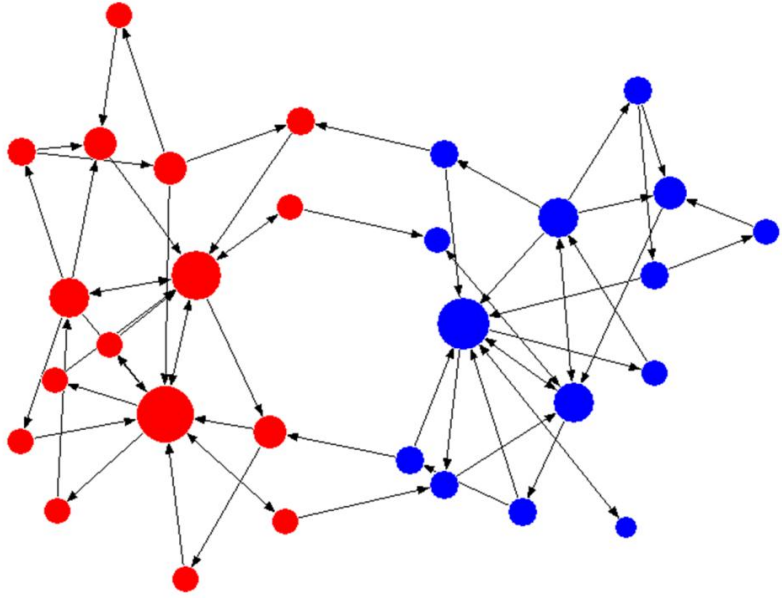

**Fig. 3** Example topology of a layered mixing network when $N_r = N_s = 0.5N, N = 30$. Blue nodes belong to the shadow bank layer. Red nodes belong to the regulated bank layer.

## 4  Result

The experimental conditions are as follows. The number of external asset classes is $M = 2$. The values of the parameters are $\theta = 0.3$, $\gamma_s = 0.06$ for shadow banks and $\gamma_r = 0.1$ for regulated banks, $\kappa = 0.05$, and $\rho = 0.25$. The number of nodes is $N = 500$ for the random mixing network. The number of nodes is $N = 500$ for the asset-correlated mixing network. The number of shadow banks is determined by $f$ in both topologies. The number of nodes is $N_r = N_s = 500$ for the layered mixing network. The fraction of shadow banks is fixed at $f = 0.5$. The creditor and debtor banks of an inter-layer interbank loan are chosen randomly.

(1) Random mixing network: The curve (a) in Fig. 4 shows the number of bank bankruptcies $F(f)$ as a function of the fraction of shadow banks. The number of bankruptcies is larger than that shown by the straight line (c) $F(f) = F(0) + fN$, which assumes every shadow bank goes bankrupt, when the fraction of shadow banks is 10% through 50%. This is a clear evidence of financial contagion from a shadow banking system to regulated banks.

(2) Asset-correlated mixing network: The number of bank bankruptcies shown by the curve (a) in Fig. 5 is much larger than that shown by the curve (b), which is the number of bankruptcies for a hypothetical network where the all banks have the average equity capital ratio $\gamma_s = \gamma_r = \bar{\gamma} = \Sigma c_n / \Sigma a_n$ uniformly. Given the total amount of capital buffer, banks having heterogeneous equity capital ratio is more vulnerable than banks having homogeneous equity capital ratio.

(3) Layered mixing network: Fig. 6 shows the ratio of increase in the number of bank bankruptcies as a function of $q$. The ratio is defined by $R(q) = [F(q) - F(0)]/F(0)$. The relative denseness $q$ is the inter-layer denseness relative to the intra-layer denseness. $F_s(0) = 468$ (94%) shadow banks and $F_r(0) = 124$ (25%) regulated banks go bankrupt when $q = 0$ and $R = 1$. 592 banks (59%) go bankrupt in total. The number of bankruptcies increases even for very small values of $q \simeq 0.05$ because more regulated banks go bankrupt (see the curve (c)). There are few surviving shadow banks regardless of the value of $q$ (see the curve (b)). This implies financial contagion from shadow banks to regulated banks. Decoupling a financial system into multiple layers does not necessarily mitigate the severity of financial contagion because $q$ cannot be zero under a practical circumstance. On the other hand, the number of bankruptcies does not increase for large values of $q \simeq 0.2$. The financial distress from shadow banks is leveled off by many neighboring regulated banks. But the number of bankruptcies is still much larger than that when the layers are decoupled completely. Note that the reason why the curves fluctuate is not evident. This is for future study.

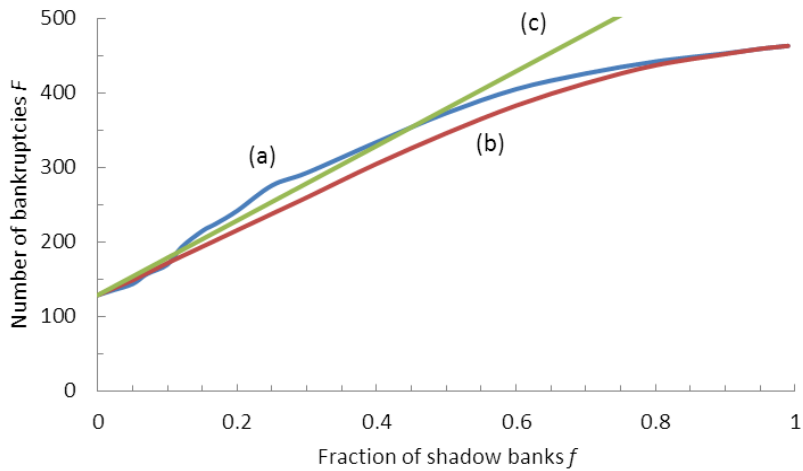

**Fig. 4** Number of bank bankruptcies as a function of the fraction of shadow banks. (a) Random mixing network, (b) hypothetical network where the all banks have the average equity capital ratio ($\bar{\gamma} = \Sigma c_n/\Sigma a_n$) uniformly, and (c) $F(f) = F(0) + fN$.

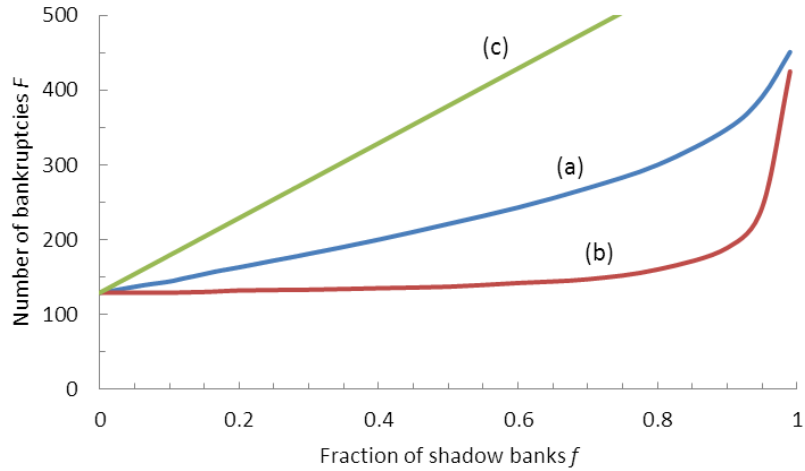

**Fig. 5** Number of bank bankruptcies as a function of the fraction of shadow banks. (a) Asset-correlated mixing network, (b) hypothetical network where every bank has the average equity capital ratio ($\bar{\gamma} = \Sigma c_n/\Sigma a_n$) uniformly, and (c) $F(f) = F(0) + fN$.

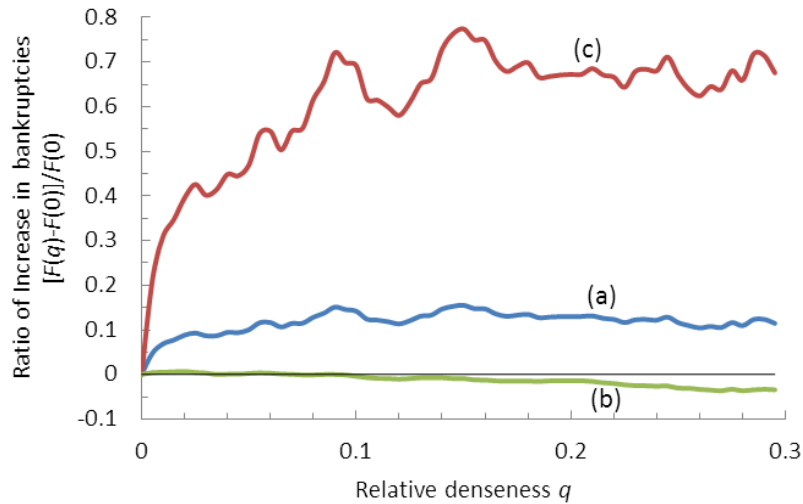

**Fig. 6** Ratio of the increase in the number of bank bankruptcies $R(q) = [F(q) - F(0)]/F(0)$ as a function of $q$. The relative denseness $q$

is the inter-layer denseness relative to the intra-layer denseness in the layered mixing network. (a) Entire financial system, (b) banks in the shadow bank layer, and (c) banks in the regulated bank layer.

## 5  Conclusion

The findings in this study include:
(1) Random mixing network: Financial contagion from shadow banks causes the bankruptcies of regulated banks when the fraction of shadow banks is 10% through 50%.
(2) Asset-correlated mixing network: The number of bankruptcies is much larger than that for a hypothetical network where every bank has the average equity capital ratio uniformly. This finding also holds true to the random mixing network.
(3) Layered mixing network: The number of bankruptcies increases even for a very small value of the inter-layer denseness relative to the intra-layer denseness.

The findings imply that failing shadow banks may affect regulated banks and consequently the entire financial system, banks having heterogeneous equity capital ratio may be vulnerable, and layer decoupling may not eliminate financial contagion. These implications are relevant to the argument to reform the capital requirement and regulatory standards.

### Acknowledgement

The authors would like to thank Hidetoshi Tanimura, Ernst&Young ShinNihon LLC, Teruyoshi Kobayashi, Kobe University, Akira Namatame, National Defense Academy, and Yuji Aruka, Chuo University for their advice and discussion.